\documentclass[11pt,document,nofootinbib,superscriptaddress,onecolumn,preprintnumbers,balancelastpage]{article}

\pdfoutput=1
\usepackage{jheppub}
\usepackage{cancel}
\usepackage[normalem]{ulem}
\usepackage{slashed}
\usepackage{mathtools}
\usepackage{amsfonts}
\usepackage{amssymb}
\usepackage{epsfig}
\usepackage{rotating}
\usepackage{url}
\usepackage{times}
\usepackage{color}
\usepackage{bm}
\usepackage{xcolor,colortbl}
\usepackage{hyperref}
\usepackage{enumitem}
\usepackage{fancyhdr}
\usepackage{anyfontsize}
\usepackage{wasysym}
\usepackage{empheq}

\def\r{\rho}

\def\m{\mu}
\def\n{\nu}

\def\r{\rho}
\def\s{\sigma}

\def\a{\alpha}
\def\b{\beta}

\newcommand{\be}{\begin{equation}}
\newcommand{\ee}{\end{equation}}
\newcommand{\bea}{\begin{eqnarray}}
\newcommand{\eea}{\end{eqnarray}}

\let\isout\sout \renewcommand{\sout}[1]{\ifmmode\text{\isout{\ensuremath{#1}}}\else\isout{#1}\fi}

\definecolor{colorTM}{rgb}{.2,.7,.2}

\newcommand{\kibitz}[2]

\begin{document}

\title{Isometries and the double copy}

\author[a]{Damien~A.~Easson,}
\author[b]{Gabriel Herczeg,}
\author[b]{Tucker Manton,}
\author[a,b]{and Max Pezzelle}

\affiliation[a]{\footnotesize Department of Physics, Arizona State University, Tempe, AZ 85287-1504, USA}
\affiliation[b]{\footnotesize Brown Theoretical Physics Center, Department of Physics, Brown University, Providence, RI 02912, USA}

\date{\today}
\abstract{
In the standard derivation of the Kerr-Schild double copy, the geodicity of the Kerr-Schild vector and the stationarity of the spacetime are presented as assumptions that are necessary for the single copy to satisfy Maxwell's equations. However, it is well known that the vacuum Einstein equations \emph{imply} that the Kerr-Schild vector is geodesic and shear-free, and that the spacetime possesses a distinguished vector field that is simultaneously a Killing vector of the full spacetime and the flat background, but need not be timelike with respect to the background metric. We show that the gauge field obtained by contracting this distinguished Killing vector with the Kerr-Schild graviton solves the vacuum Maxwell equations, and that this definition of the Kerr-Schild double copy implies the Weyl double copy when the spacetime is Petrov type D. When the Killing vector is taken to be timelike with respect to the background metric, we recover the familiar Kerr-Schild double copy, but the prescription is well defined for any vacuum Kerr-Schild spacetime and we present new examples where the Killing vector is null or spacelike. While most examples of physical interest are type D, vacuum Kerr-Schild spacetimes are generically of Petrov type II. We present a straightforward example of such a spacetime and study its double copy structure. Our results apply to real Lorentzian spacetimes as well as complex spacetimes and real spacetimes with Kleinian signature, and provide a simple correspondence between real and self-dual vacuum Kerr-Schild spacetimes. This correspondence allows us to study the double copy structure of a self-dual analog of the Kerr spacetime. We provide evidence that this spacetime may be diffeomorphic to the self-dual Taub-NUT solution.}

\maketitle

\section{Introduction} 
Our understanding of the interplay between gauge and gravity theories has progressed considerably in the past few decades. Notably, a duality called the \textit{double copy} has attracted significant interest. The double copy has its historical roots in the Kawai-Lewellen-Tye (KLT) relations \cite{Kawai:1985xq}, which provide a procedure for obtaining tree-level scattering amplitudes of closed strings from two open string amplitudes. In the field theory limit, excitations of closed and open strings correspond to gravitons and non-Abelian gauge fields respectively, suggesting that amplitudes describing spin-1 and spin-2 theories are intimately related. 

Decades later, Bern, Carrasco and Johansson introduced the notion of color-kinematics duality \cite{Bern:2008qj}, which utilizes a kinematic analog of the Jacobi identity for color factors to recast gravity tree amplitudes in terms of QCD amplitudes, shedding new light on the KLT relations, and initiating a program that has become the state of the art in perturbative gravity, with applications ranging from higher loop graviton scattering \cite{Bern:2012uf} to efficiently computing conservative potentials for compact binary systems \cite{bern2019black}. 

To briefly summarize, the Yang-Mills amplitude can be written schematically as \cite{Bern:2019prr,Monteiro:2014cda} $\mathcal{A}\sim\sum_i\frac{n_ic_i}{\text{props}}$. The functions $n_i$ are so-called kinematic numerators (functions of polarizations and momenta), $c_i$ are color factors (contracted products of gauge theory structure constants), and the `props' denominator contains all relevant propagator momenta. From gauge-invariance, it follows that the color factors satisfy the Jacobi identity $c_i+c_j+c_k=0.$
The key observation is that when the Feynman diagrams are organized such that a set of kinematic numerators, say $\tilde{n}_i$, satisfy a Jacobi identity analogously to the color factors, $\tilde{n}_i+\tilde{n}_j+\tilde{n}_k=0,$ then the $c_i$ can be replaced with the $\tilde{n}_i$ to obtain $\mathcal{M}\sim\sum_i\frac{n_i\tilde{n}_i}{\text{props}}$. This object is the graviton amplitude corresponding to the graviton scattering process at the same loop order as the Yang-Mills amplitude. Since we replaced the color factor with a copy of the kinematic numerator, $\mathcal{M}$ is referred to as the \textit{double copy} of $\mathcal{A}$, and $\mathcal{A}$ is referred to as the \textit{single copy} of $\mathcal{M}$. It can be shown that the Jacobi identity for the kinematic numerators arises due to linear diffeomorphism invariance, as expected for any genuine graviton amplitude \cite{Bern:2019prr}. We can instead strip the kinematic numerator from $\mathcal{A}$ and replace it with a second color factor $\tilde{c}_i$ (again satisfying the Jacobi identity) to obtain the gauge-invariant object $\mathcal{T}\sim\sum_i\frac{c_i\tilde{c}_i}{\text{props}}$. This describes the scattering of scalar fields $\phi^{aa'}$, which transform in the adjoint representation of two Lie algebras. The scalar fields are referred to as the \textit{zeroth copy} and are described by the cubic Lagrangian $\mathcal{L}_{\text{scalar}}=\frac{1}{2}\partial_\mu\phi^{aa'}\partial^\mu\phi_{aa'}+\frac{y}{3}f_{abc}f_{a'b'c'}\phi^{aa'}\phi^{bb'}\phi^{cc'}.$

The utility of the double copy has become increasingly clear as the program has continued to advance. It has played a central role in computing four-loop amplitudes in $\mathcal{N}=4$ super Yang-Mills and $\mathcal{N}=8$ supergravity \cite{Bern:2012uf}, and it has taken on a surprisingly prominent role in computing gravitational observables \cite{Goldberger:2017frp,Luna:2017dtq,Goldberger:2017vcg,Shen:2018ebu,Cheung:2018wkq,Kosower:2018adc,Bern:2019nnu,Antonelli:2019ytb,Bern:2019crd,Kalin:2019rwq},  which has become more important than ever due to the discovery of gravitational waves \cite{LIGOScientific:2016aoc,LIGOScientific:2017vwq}. Despite these significant advances, a fundamental theoretical understanding of the double copy remains elusive. While some efforts have been made to develop a Lagrangian formulation of the double copy \cite{Bern:2010yg, Tolotti:2013caa, Cheung:2016say}, there is much that remains obscure regarding why the double copy works, and its scope of applicability.

One approach to grappling with these shortcomings is to try to understand the double copy at the classical level. The first attempt to understand the classical aspects of the double copy was initiated by Monteiro, O'Connell and White in \cite{Monteiro:2014cda}, where it was observed that a particular subset of solutions of Einsteins equations---those of the Kerr-Schild form---could be mapped to solutions of the flat-space Maxwell equations using a procedure that closely mimics color-kinematics duality. Specifically, associated to a stationary solution of the Einstein equations that admits the decomposition

\begin{equation}
\label{Kerr-Schild-first}    g_{\mu\nu}=g^{(0)}_{\mu\nu}+\phi k_\mu k_\nu,
\end{equation}
where $g^{(0)}_{\mu\nu}$ is a background metric usually taken to be flat space and the vector $k_\mu$ is null and geodesic with respect to $g^{\mu\nu}$, and $g_{(0)}^{\mu\nu}$ is a gauge potential 
\begin{eqnarray}
    A_\mu^a=T^a\phi k_\mu
\end{eqnarray}
that is a solution to the linearized Yang-Mills equations (Maxwell theory) over $g_{\mu\nu}^{(0)}$. The color factor $T^a$ is trivial in Maxwell theory. Identifying the linear graviton as $\kappa h_{\mu\nu}=\phi k_\mu k_\nu,$ we observe that the gauge potential is obtained from the graviton by replacing a null vector $k_\mu$ with a color factor $T^a$, analogous to replacing a kinematic numerator in the graviton amplitude $\mathcal{M}$ with a color factor to obtain the Yang-Mills amplitude $\mathcal{A}$. We can further replace the second null vector with another color vector $T^{a'}$ to obtain a scalar $\phi^{aa'}=T^aT^{a'}\phi,$ which is a solution to the flat space wave equation $\Box^{(0)}\phi^{aa'}=0,$ \textit{i.e.} the linearized bi-adjoint scalar theory.\footnote{In the main text, we slightly alter the notation for the zeroth copy in order to stay consistent with our more general approach to the copying program. We will also neglect the passive color indices, as is common in the classical double copy literature.} We note that an alternative but equivalent prescription for identifying the single and zeroth copy fields from the graviton for stationary spacetimes is by simply contracting the timelike Killing vector $X=\partial_t$ as $A_\mu=h_{\mu\nu}X^\nu$, $\phi=h_{\mu\nu}X^\mu X^\nu$. Such an approach was considered for timelike and null Killing vectors for stationary and wave solutions respectively in, \textit{e.g.},  \cite{Carrillo-Gonzalez:2017iyj,Bah:2019sda}, and a special `screw' Killing vector played a significant role in the analysis in \cite{Ilderton:2018lsf}. This perspective is particularly useful for the more general prescription that we present in section \ref{New-KS-DC}, which we will elaborate on momentarily.

A complementary realization of the classical double copy was presented in \cite{Luna:2018dpt}, and is referred to as the Weyl double copy. The Weyl double copy is formulated using spinorial language where the key gravitational variable is the completely symmetric Weyl spinor $\Psi_{ABCD}$. The original construction of the Weyl double copy applies to Petrov type D and type N vacuum spacetimes, and the statement is that for a Weyl spinor associated to such a vacuum solution, the relationship
\begin{equation}\label{WeylDoubleCopy}
    \Psi_{ABCD}=\frac{1}{S}\varphi_{(AB}\varphi_{CD)}
\end{equation}
identifies a Maxwell spinor $\varphi_{AB}$ and (complex) scalar field $S$ that respectfully satisfy the vacuum Maxwell and wave equations about an appropriate flat limit of the full spacetime. The Weyl double copy has since been shown to emerge from the so-called Penrose transform using the twistor formalism \cite{White:2020sfn,Chacon:2021wbr,Chacon:2021lox,CarrilloGonzalez:2022ggn} and can be applied to higher Petrov types, and more recently, it has been applied to certain non-vacuum solutions using both the twistor \cite{Armstrong-Williams:2023ssz} and spinor \cite{Easson:2021asd,Easson:2022zoh} constructions.

Since the appearance of \cite{Monteiro:2014cda}, numerous other approaches to the classical double copy have proliferated \cite{Luna:2015paa,    White:2016jzc, Goldberger:2016iau, DeSmet:2017rve, Adamo:2017nia, Luna:2018dpt, Berman:2018hwd, Lee:2018gxc, Gurses:2018ckx, Bahjat-Abbas:2018vgo, Bah:2019sda, Cho:2019ype, Alawadhi:2019urr, Banerjee:2019saj, Kim:2019jwm, Huang:2019cja, Arkani-Hamed:2019ymq,  CarrilloGonzalez:2019gof,  Luna:2020adi, Keeler:2020rcv, Easson:2020esh, Elor:2020nqe, prabhu2020classical, alawadhi2020s, cheung2020scattering, delaCruz:2020bbn, Emond:2020lwi, White:2020sfn, 1969JMP....10.1842D, Godazgar:2020zbv, Chacon:2020fmr, Berman:2020xvs, Ben-Shahar:2021zww, Monteiro:2021ztt,  guevara2021reconstructing, Gonzalez:2021bes, Chacon:2021hfe, Cho:2021nim, Chacon:2021wbr, Easson:2021asd, Campiglia:2021srh, Farnsworth:2021wvs, guevara2021worldsheet, Adamo:2021dfg,  Godazgar:2021iae,  Kosower:2022yvp,   CarrilloGonzalez:2022mxx, Alkac:2021seh, Alkac:2021bav, Chacon:2021lox,  Chen:2021chy, Gonzalez:2021ztm, Angus:2021zhy, Moynihan:2021rwh, Alkac:2022tvc, Han:2022ubu, Mao:2021kxq, Nagy:2022xxs, Armstrong-Williams:2022apo, CarrilloGonzalez:2022ggn, Ben-Shahar:2022ixa, Didenko:2022qxq, Easson:2022zoh, Chawla:2022ogv, Armstrong-Williams:2023ssz, chawla2023black, lipstein2023self, gonzalez2023mini, Bonezzi:2023pox, Brown:2023zxm, Borsten:2023paw, Ball:2023xnr}
---however, the main purpose of this work is to revisit the original Kerr-Schild double copy of \cite{Monteiro:2014cda} with the goal of clarifying some technical points that have been overlooked up until now. In particular, the geodicity of the Kerr-Schild vector and the existence of a timelike Killing vector were treated as assumptions in the original presentation of the Kerr-Schild double copy. However, the geodicity of the Kerr-Schild vector and the existence of a Killing vector are \emph{consequences} of the vacuum conditions \cite{Stephani:2003tm,Bini:2010hrs}. This Killing vector can be spacelike, timelike or null with respect to the background flat metric, and the timelike character of the Killing vector is not necessary to define the double copy.  This leads us to propose a modified definition of the Kerr-Schild double copy for four-dimensional vacuum spacetimes that applies to all real vacuum spacetimes with Lorentzian signature\footnote{Technically, our definition only applies to spacetimes with an expanding Kerr-Schild vector, a caveat that excludes PP wave spacetimes.} as well as a subset of complex vacuum spacetimes and real spacetimes with Kleinian signature within the Kerr-Schild class. Moreover, the modified definition of the Kerr-Schild double copy that we present implies the Weyl double copy \cite{Luna:2018dpt} when either the self-dual or anti-self-dual part of the Weyl tensor is type D. The modified definition we propose is very simple: denoting the distinguished Killing vector associated with any vacuum Kerr-Schild spacetime by $X^\mu$ and the graviton by $h_{\mu\nu} := g_{\mu\nu} - \eta_{\mu\nu}$, the proposed single copy gauge field and zeroth copy scalar field are $A_\mu = h_{\mu\nu}X^\nu$ and $\phi = h_{\mu\nu}X^\mu X^\nu$. 

Vacuum Kerr-Schild spacetimes can be characterized very conveniently in terms of a suitable null tetrad using the spin coefficient formalism of Newman and Penrose. Since this formalism may not be entirely familiar to our audience, we devote section 
\S \ref{sec:spincoeffformalism} to a brief review of this material with a particular emphasis on the role played by differential forms. In section \S \ref{VKS} we apply this formalism to vacuum Kerr-Schild spacetimes, and in section \S \ref{New-KS-DC} we present our proposal for defining the Kerr-Schild double copy for vacuum spacetimes in terms of the distinguished null vector admitted by any such spacetime, and show that this proposal leads to single and zeroth copies that satisfy the vacuum Maxwell equations and wave equation respectively. In section \S \ref{WeylDC} we show that our proposal for the Kerr-Schild double copy for vacuum spacetimes implies the Weyl double copy when the left or right Weyl spinor is Petrov type D. Finally, in  section \S \ref{egs} we provide a number of 
examples of spacetimes that can be studied with our approach, including spacetimes with Killing vectors that are spacelike or null with respect to the background flat metric, self-dual spacetimes, and a simple, explicit example with Petrov type II.

\section{Spin coefficient formalism}\label{sec:spincoeffformalism}
Our approach to expanding on the Kerr-Schild double copy will make use of a version of the spin coefficient formalism that emphasizes the role of differential forms \cite{mcintosh1985complex}. It is somewhat regrettable that this formalism seems to be the one that is best suited for the problems we wish to address, since we expect that it is not well known to many of the readers who may be most interested in our results. Nevertheless, we feel obliged to employ it for three compelling reasons: 
\begin{enumerate}
\item 
It is the only formalism that admits an explicit procedure for constructing a broad class of vacuum Kerr-Schild spacetimes.
\item 
 It provides a very natural language for discussing the connection between the Kerr-Schild and Weyl double copies.
 \item It provides a new framework for studying self-dual solutions in the context of the double copy.
\end{enumerate}
 Throughout the article, we will restrict our attention to four-dimensional spacetimes. However, we will allow spacetime to be complex in general, and we make use of the convention that quantities decorated with a tilde should be understood as the complex conjugate of the corresponding quantity when the spacetime is real and Lorentzian, and otherwise should be regarded as independent. We will refer to quantities without tildes as having \emph{left} chirality, while quantities with tildes will be said to have \emph{right} chirality. The primary object of study in this formalism is a null co-tetrad $\theta^a$, where $a$ runs from 1 to 4. The metric can be recovered from the co-tetrad as 
\be  
ds^2 = 2(\theta^1\theta^2 - \theta^3\theta^4).
\ee  
The co-tetrad is related to the standard tetrad $\lambda_a = (l,n,m,\tilde{m})$ by $\theta^a_\mu\lambda_b^\mu = \delta^a_b$, and obeys
\be  
\theta^1_\mu = n_\mu, \qquad \theta^2_\mu = l_\mu, \qquad \theta^3_\mu = -\tilde{m}_\mu, \qquad \theta^4_\mu = \tilde\theta^3_\mu = -m_\mu.
\ee  
We also define the directional derivatives
\be  
D = l^\mu\partial_\mu, \qquad \Delta = n^\mu\partial_\mu, \qquad \delta = m^\mu\partial_\mu, \qquad \tilde\delta = \tilde{m}^\mu\partial_\mu. \label{directional}
\ee  
The exterior derivative can be expressed in terms of the co-tetrad and directional derivatives as
\be  \label{d}
d = \theta^1D + \theta^2\Delta + \theta^3\delta + \theta^4\tilde\delta.
\ee
It is often convenient to expand differential forms in terms of wedge products of the co-tetrad one-forms, for which we use the compact notation
\be  
\theta^{ab...d} \equiv \theta^a\wedge\theta^b\wedge\ldots\wedge\theta^d.
\ee  
It is also often useful to decompose two-forms into self-dual and anti-self-dual parts. For this, we define the following basis of two-forms:
\be \label{self-dual}
Z^0 =\theta^{13}, \qquad Z^1  = \theta^{12} - \theta^{34}, \qquad Z^2  =  \theta^{42},
\ee 
\be 
\tilde{Z}^0 =\theta^{14}, \qquad \tilde{Z}^1  = \theta^{12} + \theta^{34}, \qquad \tilde{Z}^2  =  \theta^{32},
\ee
where $Z^j$ are self-dual and $\tilde{Z}^j$ are anti-self-dual---i.e. $*Z^j = iZ^j$ and $*\tilde{Z}^j = -i\tilde{Z}^j$.\footnote{Note that this self-duality condition is appropriate for real spacetimes with Lorentzian signature. For other cases, this condition must be modified accordingly.} 

Now the spin coefficients can be computed either from the first Cartan structure equations
\bea 
d\theta^1 &=& (\gamma + \tilde\gamma)\theta^{12} + (\tilde\alpha + \beta - \tilde\pi)\theta^{13} + (\alpha + \tilde\beta - \pi)\theta^{14}   - \tilde\nu\theta^{23} - \nu\theta^{24} - (\mu - \tilde\mu)\theta^{34} ~~~~~~~~ \nonumber \\
d\theta^2 &=& (\epsilon + \tilde\epsilon)\theta^{12}  + (\tau - \tilde\alpha - \beta)\theta^{23} + (\tilde\tau -\alpha - \tilde\beta)\theta^{24} + \kappa\theta^{13} + \tilde\kappa\theta^{14}   - (\rho - \tilde\rho)\theta^{34} ~~~~~~~~ \nonumber \\
d\theta^3 &=& -(\tilde\tau + \pi)\theta^{12} - (\tilde\rho + \epsilon - \tilde\epsilon)\theta^{13} - \tilde\sigma\theta^{14}   + (\mu + \tilde\gamma - \gamma )\theta^{23} + \lambda\theta^{24} + (\alpha - \tilde\beta)\theta^{34} ~~~~~~~~ \nonumber  \\
d\theta^4 &=& -(\tau + \tilde\pi)\theta^{12} - (\rho + \tilde\epsilon - \epsilon)\theta^{14} - \sigma\theta^{13}   + (\tilde\mu + \gamma - \tilde\gamma )\theta^{24} + \tilde\lambda\theta^{23}  -(\tilde\alpha - \beta)\theta^{34} ~~~~~~~~ \label{firstCartan}
\eea 
or from the the bivector equations\footnote{An earlier version of this article contained a number of typographical errors in equations \eqref{firstCartan}-\eqref{bivector}. We are grateful to Michael Graesser for bringing this to our attention.}
\bea
dZ^0 &=& (2\gamma - \mu)\theta^{123} -\lambda\theta^{124} + (\pi - 2\alpha)\theta^{134} + \nu\theta^{234}  \nonumber \\
dZ^1 &=& -2\tau\theta^{123} + 2\pi\theta^{124} +2\rho\theta^{134} - 2\mu\theta^{234}  \nonumber \\
dZ^2 &=& \sigma\theta^{123} + (\rho - 2\epsilon)\theta^{124} - \kappa\theta^{134} + (2\beta - \tau)\theta^{234},  \label{bivector} \eea  
along with the corresponding expressions for $d\tilde{Z}^j$. For real spacetimes, the bivector equations for the self-dual two-forms $Z^j$ are sufficient to determine all of the spin coefficients, while for complex spacetimes the bivector equations conveniently decouple the left and right spin coefficients. 

The Newman-Penrose equations, which encode the dynamics of general relativity in the spin coefficient formalism, and the Bianchi identities can also be cast in terms of equations involving bivectors and spin coefficients, but since we will not use them, we refer the reader to \cite{mcintosh1985complex} for details. The Newman-Penrose equations and Bianchi identities can be found in \cite{newman1962approach,Stephani:2003tm} and many other places, so we will not record them here. We will now examine the application of this formalism to a broad class of vacuum Kerr-Schild spacetimes.

\section{Vacuum Kerr-Schild spacetimes}\label{VKS}
We now direct our attention to Kerr-Schild spacetimes, which are characterized by a metric that can be expressed in the following form
\be
g_{\mu\nu} = g^{(0)}_{\mu\nu} + 2V l_{\mu}l_{\nu}, \label{Kerr-Schild}
\ee
where $g^{(0)}_{\mu\nu}$ is a flat metric, $V$ is a scalar function, and $l_\mu = \theta^2_\mu$ is null, and expanding.\footnote{In terms of spin coefficients, the condition for $l_\mu$ to be geodesic is $\kappa = \tilde\kappa = 0$, the shear-free condition is $\sigma = \tilde\sigma = 0$, and the expanding condition is $\rho + \tilde\rho \neq 0$. We also note that we have employed a slightly different notation for the Kerr-Schild form in \eqref{Kerr-Schild}compared with \eqref{Kerr-Schild-first}, because here we use a different scaling for the Kerr-Schild vector than the one typically used in the Kerr-Schild double copy.} The vacuum Einstein equations then imply that $l_\mu$ is also geodesic and shear-free with respect to both the full metric and the background flat metric (see theorem (32.1) of \cite{Stephani:2003tm} and Chapter 6 section 57 of \cite{Chandrasekhar:1985kt}). Note that $l_\mu$ is null with respect to $g_{\mu\nu}$ if and only if it is null with respect to $g^{(0)}_{\mu\nu}$. Much of what appears in this section can be found in \cite{mcintosh1988single,mcintosh1989kerr} as well as chapter 32 of \cite{Stephani:2003tm}.
A co-tetrad corresponding to the metric \eqref{Kerr-Schild} is given by
\begin{eqnarray}
\theta^1 &=& dv + V\theta^2 \nonumber\\
\theta^2 &=&  du+ \tilde{Y}d\zeta + Y d\tilde{\zeta} + Y\tilde{Y}dv  \nonumber\\
\theta^3 &=&  d\tilde\zeta + \tilde{Y}dv  \nonumber \\
\theta^4 &=& d\zeta + Ydv.
\label{eq:co-tetrad}
\end{eqnarray}
where $Y$, $\tilde{Y}$ and $V$ are generically complex functions, and all of the coordinates $u$, $v$, $\zeta$, $\tilde\zeta$ are generically complex and independent. When the spacetime is real, $Y$ and $\tilde{Y}$ are complex conjugates and V is real; $u,v$ are real light-cone coordinates, and $\zeta,\tilde{\zeta}$ are complex conjugate coordinates related to the usual Cartesian coordinates $(t,x,y,z)$ by 
\be
\label{eq:coor}
u = \tfrac{1}{\sqrt{2}}(t - z), \quad v = \tfrac{1}{\sqrt{2}}(t+ z), \quad \zeta = \tfrac{1}{\sqrt{2}}(x + iy).
\ee
In these coordinates, the flat metric appearing in \eqref{Kerr-Schild} is
\be  
ds_0^2 = g^{(0)}_{\mu\nu}dx^\mu dx^\nu = 2(dudv - d\zeta d\tilde\zeta).
\ee 
The directional derivatives associated with \eqref{eq:co-tetrad} are
\bea
D &=& \partial_v - Y\partial_\zeta - \tilde{Y}\partial_{\tilde{\zeta}} + Y\tilde{Y}\partial_u\,\nonumber \\
\Delta &=& \partial_u - VD\,\nonumber\\
\delta &=& \partial_{\tilde{\zeta}} - Y\partial_u\, \nonumber \\
\tilde{\delta} &=& \partial_{\zeta} - \tilde{Y}\partial_u.
\label{tetrad vectors}
\eea
The conditions that $l_\mu = \theta^2_\mu$ be geodesic and shear-free are encoded in the vanishing of the spin coefficients $\kappa = DY$, $\sigma = \delta Y$, $\tilde\kappa = D\tilde{Y}$, and $\tilde\sigma = \tilde\delta\tilde{Y}$, with the latter two conditions following from the first two for real spacetimes. Note that because $\kappa$ and $\sigma$ do not depend on $V$, $l_\mu$ is shear-free and geodesic with respect to the full spacetime if and only if it is shear-free and geodesic with respect to the background flat spacetime. Setting $\kappa = \sigma = 0$ is then equivalent to the following pair of partial differential equations for $Y$
\be
Y_{,v} = YY_{,\zeta}\,, \qquad Y_{,\tilde{\zeta}} = YY_{,u}\label{PDEs}\,,
\ee
with analogous expressions for $\tilde{Y}$ following from $\tilde{\kappa} = \tilde{\sigma} = 0$. Note that the pair of equations above also implies that $Y$ is a solution of the flat-space wave equation $\square^{(0)} Y = 0$, where $\square^{(0)} = 2(\partial_u\partial_v - \partial_\zeta\partial_{\tilde{\zeta}})$. The general solution to these equations is found using Kerr's theorem \cite{Penrose:1967wn, cox1976conventional} which characterizes the most general shear-free, null geodesic congruence on Minkowksi space by specifying $Y$ implicitly as the solution of the zero locus of a homogeneous function of twistor variables. Alternatively, equations \eqref{PDEs} can be solved via the method of characteristics \cite{john1982partial}. In either case, one finds that $Y$ solves \eqref{PDEs} if and only if it is a solution of an \emph{algebraic} equation
\be  
f(Y, u + Y\tilde{\zeta}, \zeta + Yv) = 0 \label{zero set}
\ee 
for an arbitrary analytic function $f$. Having imposed $\kappa = \sigma = 0$, the remaining
non-vanishing left spin coefficients are 
\bea \label{spinCoef}
\rho &=& \tilde\delta Y, \,\,\,\,\,\qquad \tau = \Delta Y, \,\,\,\,\,\qquad \mu = V\rho, \nonumber \\
\gamma &=& \frac{1}{2}\left[DV + V(\rho - \tilde\rho)\right], \qquad \nu = \tilde\delta V - V\tilde\tau,
\eea 
with analogous expressions holding for the right spin coefficients, as always. 

Now the vacuum Newman-Penrose equations can be organized into three groups \cite{herlt1980kerr, Stephani:2003tm}. The first group
\be
D\rho = \rho^2, \quad D\tau = \tau\rho, \quad \delta\tau = \tau^2, \quad \delta\rho = (\rho - \tilde\rho)\tau, \quad \tilde\delta\tau - \Delta\rho = \tau\tilde\tau + V\rho^2, \label{NPgroup1} 
\ee 
is satisfied automatically after taking into account commutator identities for the directional derivatives \eqref{directional}. The second group 
\begin{gather} 
\delta\gamma = \tau(\mu + \gamma) \nonumber \\
D\gamma = \mu\rho + \gamma(\rho -\tilde\rho) \nonumber \\
D\mu = \mu(\rho + \tilde\rho) + \gamma(\rho -\tilde\rho) \nonumber \\
\delta\nu -\Delta\mu = \mu^2 + (\gamma + \tilde\gamma)\mu + \tau\nu \label{NPgroup2}
\end{gather} 
 remain to be solved, and the third group determines the non-vanishing Weyl scalars in terms of the spin coefficients: 
\bea  
\Psi_2 &=& \mu\rho + \gamma(\rho -\tilde\rho) \nonumber \\ 
\Psi_3 &=& \tilde\delta\gamma + \rho\nu - \tilde\tau\gamma \nonumber \\
 &=& D\nu - \tilde\tau\mu \nonumber \\ &=& -\tilde\delta\mu + (\rho -\tilde\rho)\nu\nonumber \\
\Psi_4 &=& \tilde\delta\nu - \tilde\tau\nu. \label{NPgroup3}
\eea 
The vacuum conditions solving \eqref{NPgroup2} were provided by Debney, Kerr and Schild in \cite{1969JMP....10.1842D}. First, a particular form for the analytic function in \eqref{zero set} is required:
\be \label{fvac}
f(Y, u + Y\tilde{\zeta}, \zeta + Yv) = \Phi(Y) + (\tilde{c}Y + a)(\zeta + Yv) - (bY + c)(u + Y\tilde{\zeta}) 
\ee
where $\Phi(Y)$ is an arbitrary function, and $a,b,c,\tilde{c}$ are constants. Second, the function $V$ is fixed as
\be  
V = \frac{m}{2P^3}(\rho + \tilde\rho), \label{Vvac}
\ee
where 
\be 
P = a + \tilde{c}Y + c\tilde{Y} + bY\tilde{Y}. 
\ee 
Note that $DY = D\tilde{Y} = 0$ implies that $DP = 0$. Now the vacuum condition \eqref{Vvac}, together with the Newman-Penrose equation $D\rho = \rho^2$ from \eqref{NPgroup1} imply that 
\be 
\gamma = \frac{m\rho^2}{2P^3}. \label{gammaVac}
\ee 
Acting with $\Delta$ and $\tilde{\delta}$ on \eqref{zero set} with $f$ given by \eqref{fvac} lead to, respectively 
\be
\tau = \frac{bY + c}{f_Y}, \qquad \rho = -\frac{P}{f_Y}, \label{rhotauVac1}
\ee 
and eliminating $f_Y$ leads to 
\be 
\tau = -\frac{\rho}{P}(bY + c). \label{rhotauVac2}
\ee 
Now consider the vector field
\be 
X = a\partial_u + b\partial_v + c\partial_\zeta + \tilde{c}\partial_{\tilde{\zeta}}. \label{Killing}
\ee 
Acting with $X$ on \eqref{zero set} with $f$ given by \eqref{fvac} leads to $X(Y) = 0$
which together with the right-handed analog $X(\tilde{Y}) = 0$ implies that $X(P) = 0$. Also, note that $X(Y) = X(\tilde{Y}) = 0$ implies $[X,\delta] = [X,\tilde\delta] = 0$, which in turn implies that $X(\rho) = X(\tilde\rho) = 0$. Now, acting with $X$ on \eqref{Vvac} results in $X(V) = 0$. Finally, since $X$ is clearly a Killing vector for the background metric $g^{(0)}_{\mu\nu}$, and $X(Y) = X(\tilde{Y}) = 0$ implies $\mathcal{L}_X\theta^2 = 0$, we conclude that $X$ is a Killing vector for the full spacetime.

\section{Kerr-Schild double copy from the distinguished Killing vector $X$}\label{New-KS-DC}

Here we show that the Kerr-Schild double copy can be defined for any vacuum Kerr-Schild spacetime with an expanding and shear-free Kerr-Schild vector. Denoting the Kerr-Schild graviton by 
$$
h_{\mu\nu} = 2Vl_\mu l_\nu,
$$
the familiar recipe for the Kerr-Schild double copy can be expressed as 
$$
A_\mu = h_{\mu\nu} X^\nu
$$
where $X = \frac{\partial}{\partial t}$ is a timelike Killing vector. Here we make the same ansatz for the distinguished Killing vector $X$ defined by \eqref{Killing}, which leads to
\be 
A_\mu = 2PV l_\mu. \label{gauge}
\ee 
We now show that this is indeed a solution of the flat-space vacuum Maxwell equations when the vacuum Newman-Penrose equations \eqref{NPgroup1}, \eqref{NPgroup2}, \eqref{NPgroup3} are satisfied. The proof is somewhat technical, and the reader who is only interested in making use of our results may wish to proceed directly to section \ref{WeylDoubleCopy}. However, since it is a core result, we present the details of most calculations explicitly in this section, relegating only the most tedious steps to appendix \ref{SomeCalcs}. To begin, we calculate the field strength $F \equiv dA$:
\bea 
F= \varphi_i Z^i + \tilde{\varphi}_i\tilde{Z}^i
\eea 
where  
\be \label{NPMaxwell}  
\varphi_0 = 0, \qquad \varphi_1 = 2P\gamma, \qquad \varphi_2 =  2P \nu + 2V\tilde{\delta}P,
\ee 
and where we used equation \eqref{d} and the first Cartan structure equations \eqref{firstCartan} to evaluate the exterior derivatives, and equations \eqref{spinCoef} and \eqref{self-dual} to simplify the result.  Taking the self-dual projection $\mathcal{F} \equiv \frac{1}{2}(F - i*F)$ leads to 
\be  
\mathcal{F} = \varphi_1 Z^1 + \varphi_2 Z^2.
\ee  
It's worth noting that we get the same result for the self-dual projection whether we use the Hodge star associated with the full spacetime or the one associated with the background flat spacetime. Hence, $F$ is a solution of the vacuum Maxwell equations on the full spacetime, if and only if  it is a solution on the background spacetime if and only if $d\mathcal{F} = 0$. Strictly speaking, at this stage we are done. We know that $X_\mu = g_{\mu\nu}X^\nu$ is a solution of Maxwell's equations on the spacetime with metric $g_{\mu\nu}$ whenever $X^\nu$ is a Killing vector for $g_{\mu\nu}$. On the other hand, $A_\mu = h_{\mu\nu}X^\nu$ is gauge equivalent to $X_\mu$, so it must also be a solution of Maxwell's equations on the full spacetime with metric $g_{\mu\nu}$ and hence, it must also be a solution on the background flat spacetime. Nevertheless, it's a worthwhile exercise to show how $d\mathcal{F} = 0$ follows from the Newman-Penrose equations \cite{newman1962approach} by carrying out an explicit calculation.

Since the field strength and its self-dual projection are written naturally in terms of spin coefficients and self-dual forms on the full spacetime, we will use the tetrad for the full spacetime rather than the background spacetime for our calculations. One can compute the exterior derivative in this basis using \eqref{d} and \eqref{bivector} to find 
\bea  
\frac{1}{2}d\mathcal{F} &\equiv& J \nonumber \\
&=& J_{123}\theta^{123} + J_{124}\theta^{124} + J_{134}\theta^{134} + J_{234}\theta^{234},
\eea  
where the three-form current components are
\bea 
J_{123} &=& \delta(P\gamma) - 2\tau P \gamma \\
J_{124} &=& \tilde{\delta}(P\gamma) - D(P\nu + V\tilde{\delta}P) + \rho(P\nu + V\tilde{\delta}P) \\
J_{134} &=& -D(P\gamma) + 2\rho P \gamma \\
J_{234} &=& -\Delta(P\gamma) - 2V\rho P \gamma + \delta(P\nu + V\tilde{\delta}P) - \tau(P\nu + V\tilde{\delta}P).
\eea  
Let's look at this component by component, starting with $J_{123}$. We have
\be 
J_{123} = \gamma\delta P + P\delta\gamma - 2\tau P \gamma.
\ee 
Using the shear-free and geodesic conditions $\kappa = \sigma = 0 \iff DY = \delta Y = 0$, we have $\delta P = \tilde{\rho}(bY + c)$, and from the Newman-Penrose equations \eqref{NPgroup2}, we have $\delta\gamma = \tau(\gamma + \mu) = \tau(\gamma + \rho V)$. Using these equations along with \eqref{rhotauVac2} 
\be 
J_{123} =(bY + c)\left[ \gamma(\rho + \tilde\rho) - \rho^2V \right],
\ee 
which vanishes in light of \eqref{gammaVac}.
 The next simplest component of $J$ is
\be
J_{134} = -PD\gamma + 2\rho P \gamma,
\ee 
where we used that $DY = D\tilde{Y} = 0 \implies DP = 0$. Now, from the Newman-Penrose equations \eqref{NPgroup2} we have 
\bea   
D\gamma &=& \mu\rho + \gamma(\rho -\tilde\rho) \\
&=& \rho^2 V + \gamma(\rho -\tilde\rho),
\eea 
so that 
\be
 J_{134}
 = P\left[ \gamma(\rho + \tilde\rho) - \rho^2 V \right]
\ee
which once again vanishes taking into account \eqref{gammaVac}.

To compute $J_{124}$, 
we note that from \eqref{NPgroup1} we have the Newman-Penrose equation $D\rho = \rho^2$, which implies that $D\tilde\delta P = \rho\tilde\delta P$.
We also use two of the Newman-Penrose equations \eqref{NPgroup3}
\bea  
\tilde\delta\gamma &=& -\rho\nu + \tilde\tau\gamma + \Psi_3 \\
D\nu &=& \tilde{\tau}\mu + \Psi_3 , \nonumber
\eea 
and the fact that the expression for $\gamma$ in \eqref{spinCoef} implies $\gamma - DV = -\tilde\gamma$.

Now we compute
\bea 
J_{124} &=& \tilde\delta(P\gamma) - PD\nu - \tilde\delta P DV - VD\tilde\delta P + \rho(P\nu + V\tilde\delta P) \nonumber \\
&=& P\tilde\delta\gamma + \gamma\tilde\delta P - PD\nu - \tilde\delta P DV - V\rho\tilde\delta P + \rho P \nu + \rho V \tilde\delta P \nonumber \\
&=& P\tilde\tau(\gamma - \rho V) - \tilde\delta P\tilde\gamma \nonumber \\
&=& (b\tilde{Y} + \tilde{c})(\rho\tilde\rho V - \tilde\rho\gamma - \rho\tilde\gamma)
\eea 
which once again vanishes by virtue of \eqref{gammaVac}.

The calculation for the last component is a bit more involved. We have
{\small
\bea 
J_{234}
&=& -\Delta(P\gamma) - 2V\rho P \gamma + \delta(P\nu + V\tilde{\delta}P) - \tau(P\nu + V\tilde{\delta}P) \nonumber \\
&=& P\Delta(\mu - \gamma) - \gamma\Delta P - 2V\rho P \gamma + \delta P \nu + P(\mu^2 + (\gamma + \tilde\gamma)\mu) + \delta V\tilde\delta P + V\delta\tilde\delta P - \tau V\tilde\delta P \nonumber \\
&=& P\Delta\left(\frac{m}{2P^3}\rho\tilde\rho\right) - \gamma\Delta P + \delta P \nu + P\rho\tilde\rho V^2 + \tilde\nu\tilde\delta P + V\delta\tilde\delta P \nonumber \\
&=& P\Delta\left(\frac{m}{2P^3}\rho\tilde\rho\right) + |bY + c|^2\frac{\rho\tilde\rho}{P}V + (bY + c)\tilde\rho\nu + P\rho\tilde\rho V^2 + \tilde\nu(b\tilde{Y} + \tilde{c})\rho + Vb\rho\tilde\rho, \label{J234calc}
\eea 
}
where we have defined $|bY + c|^2 \equiv (bY + c)(b\tilde{Y} + \tilde{c})$. In going from the first line to the second line, we have used the vacuum Newman-Penrose equation 
\be  
\delta\nu - \Delta\mu = \mu^2 + (\gamma +\tilde\gamma)\mu + \tau\nu.
\ee
In going from the second line to the third, we have used $\mu = \rho V$ and the right-handed analog of $\nu = \tilde\delta V - V\tilde\tau$ from \eqref{spinCoef}, as well as \eqref{Vvac} and \eqref{gammaVac}. In going from the fourth line to the fifth, we evaluated all of the derivatives of $P$ except those in the first term, which involved making use of the Newman-Penrose equation $\delta\rho = (\rho - \tilde\rho)\tau$ from \eqref{NPgroup1}, and we have simplified the result using \eqref{Vvac}, \eqref{gammaVac} and \eqref{rhotauVac2} . We now turn our attention to the very first term:
{\small
\bea  
P\Delta\left(\frac{m}{2P^3}\rho\tilde\rho\right) 
&=& \frac{m}{2P^3}\left[-3\rho\tilde\rho\Delta P + P(\tilde\rho\Delta\rho + \rho\Delta\tilde\rho)\right] \nonumber \\
&=& -3\frac{V}{P}|bY + c|^2\rho\tilde\rho \nonumber \\
&\phantom{=}& - \tilde{\rho}(bY + c)\nu + \frac{m}{P^4}|bY + c|^2\rho\tilde\rho^2 - b\gamma\tilde\rho - \frac{m}{2P^2}V\tilde\rho\rho^2 \nonumber \\
&\phantom{=}& - \rho(b\tilde{Y} + \tilde{c})\tilde\nu + \frac{m}{P^4}|bY + c|^2\tilde\rho\rho^2 - b\tilde\gamma\rho - \frac{m}{2P^2}V\rho\tilde\rho^2 \nonumber \nonumber \\
&=& -\frac{V}{P}|bY + c|^2\rho\tilde\rho - \tilde{\rho}(bY + c)\nu - \rho(b\tilde{Y} + \tilde{c})\tilde\nu - b\rho\tilde\rho V - \rho\tilde\rho PV^2. \label{FirstTermOfJ234calc}
\eea 
}
In going from the first line to the second we have made repeated use of \eqref{Vvac}, \eqref{gammaVac} and \eqref{rhotauVac2}, as well as the equation \eqref{DeltaRhoCalc} for $\Delta\rho$ which is derived in appendix \ref{SomeCalcs} from a Bianchi identity and some of the Newman-Penrose equations. In going from the second line to the third, we have again made use of \eqref{Vvac} and \eqref{gammaVac}.
Substituting the last line of \eqref{FirstTermOfJ234calc} into the last line of \eqref{J234calc}, we see that $J_{234} = 0$. This concludes the proof that the vacuum Newman-Penrose equations imply that the gauge field \eqref{gauge} is a solution of the flat space vacuum Maxwell equations.

We may further show that the zeroth copy defined by $\phi = h_{\mu\nu}X^\mu X^\nu$ satisfies the flat-space vacuum wave equation $\square^{(0)}\phi = 0$. First, we note that the Killing vector $X$ is parallel with respect to the background flat metric, i.e $\nabla^{(0)}_\mu X^\nu = 0$. Now we calculate:
\bea 
\square^{(0)}\phi &=& \square^{(0)}(h_{\mu\nu}X^\mu X^\nu) \nonumber \\
&=& X^\mu\nabla^{(0)\nu}\nabla^{(0)}_\nu A_\mu \nonumber \\
&=& X^\mu\nabla^{(0)\nu}\nabla^{(0)}_\mu A_\nu \nonumber \\
&=& \mathcal{L}_X\big(\nabla^{(0)\nu} A_\nu\big) \nonumber \\
&=& g_{(0)}^{\mu\nu}\mathcal{L}_X\big(\nabla^{(0)}_{(\mu} A_{\nu)}\big) \nonumber \\
&=& g_{(0)}^{\mu\nu}\mathcal{L}_X\big(\tfrac{1}{2}\mathcal{L}_A\hspace{.3mm} g^{(0)}_{\mu\nu}\big) \nonumber \\
&=& \tfrac{1}{2}g_{(0)}^{\mu\nu}\big(\mathcal{L}_{[X,A]}\hspace{.3mm}g^{(0)}_{\mu\nu} + \mathcal{L}_A\mathcal{L}_X \hspace{.3mm}g^{(0)}_{\mu\nu}\big) = 0.
\eea 
In the above, we used the background flat metric to raise and lower indices. In going from the second line to the third line, we used the fact that $A_\mu$ satisfies the vacuum Maxwell equations, and in the last line we used $[X,A]^\mu = \mathcal{L}_X(g_{(0)}^{\mu\nu}A_\nu) = 0$.

\section{Weyl Double copy}\label{WeylDC}
The Weyl double copy is typically written in spinorial form as 
\be \label{SpinorWeyl}
\Psi_{ABCD} = \frac{1}{S}\varphi_{(AB}\varphi_{CD)}
\ee 
where $\Psi_{ABCD}$ is the Weyl spinor, $\varphi_{AB}$ is the field-strength spinor, and $S$ is a scalar satisfying the flat-space, vacuum wave equation $\square^{(0)}S = 0$. In terms of the Newman-Penrose Weyl scalars $\Psi_I$ and Maxwell scalars $\varphi_i$, \eqref{SpinorWeyl} becomes
\be \label{NPWeyl}
\Psi_0 = \tfrac{1}{S}\varphi_0^2, \quad \Psi_1 = \tfrac{1}{S}\varphi_0\varphi_1, \quad \Psi_2 = \tfrac{1}{3S}\left(2\varphi_1^2 + \varphi_0\varphi_2\right), \quad \Psi_3 = \tfrac{1}{S}\varphi_1\varphi_2, \quad \Psi_4 = \tfrac{1}{S}\varphi_2^2. 
\ee 
Here we show that our proposal \eqref{gauge} for the Kerr-Schild double copy obeys the Weyl double copy \eqref{NPWeyl} when the spacetime is type D. Since $\varphi_0 = \Psi_0 = \Psi_1 = 0$, the first two equations in \eqref{NPWeyl} are satisfied automatically. The first non-trivial equation is $\Psi_2 = \frac{2}{3S}\varphi_1^2$. Using $\Psi_2 = \mu\rho + \gamma(\rho -\tilde\rho)$ from \eqref{NPgroup3}, $\mu = \rho V$ from \eqref{spinCoef}, along with \eqref{Vvac} and \eqref{gammaVac} leads to $\Psi_2 = m\frac{\rho^3}{P^3}$. On the other hand, from \eqref{NPMaxwell} we have $\varphi_1 = 2P\gamma$ and using \eqref{gammaVac} leads to $\varphi_1 = \frac{m\rho^2}{P^2}.$ Using these expressions for $\Psi_2$ and $\varphi_1$ leads to 
\be \label{Seq}
S = \frac{2m\rho}{3P}.
\ee 
We now check that $S$ satisfies the flat-space wave equation, which is entirely plausible, since we have already shown that the zeroth copy $\phi$ is harmonic, and we now see that $\phi = \frac{3}{4}(S + \tilde{S})$. In order to show that $S$ is harmonic, we note from \eqref{rhotauVac2} that $\frac{\rho}{P} = -\frac{\tau}{bY + c}$. Since $DY = 0$, we write $\tau = \Delta Y = (\partial_u - VD)Y = Y_{,u}$. Now we have $\frac{\rho}{P} = -\frac{1}{b}\partial_u\log(bY + c)$, and we calculate
\bea 
\square^{(0)} S &=& -\frac{4m}{3b}\partial_u(\partial_u\partial_v - \partial_\zeta\partial_{\tilde{\zeta}})\log(bY + c) \\
&=& -\frac{4m}{3b}\partial_u\left[ \frac{b(Y_{,uv} - Y_{,\zeta\tilde{\zeta}})}{bY + c} - \frac{b^2(Y_{,u}Y_{,v} - Y_{,\zeta}Y_{,\tilde{\zeta}})}{(bY + c)^2} \right] = 0,
\eea 
where in the second line we used equations \eqref{PDEs} to eliminate the second term and their corollary $\square^{(0)} Y = 0$ to eliminate the first. 

We now turn our attention to $\Psi_3$, which can be expressed most simply as $\Psi_3 = \tilde{\delta}\gamma + \rho\nu - \tilde\tau\gamma$ from the 
Newman-Penrose equation \eqref{NPgroup3}. Using the vacuum condition $\gamma = \frac{m\rho^2}{2P^3}$ from \eqref{gammaVac}, this becomes
\be \label{Psi3EQ1}
\Psi_3 = -3\gamma\frac{\tilde\delta P}{P} + 2\gamma\frac{\tilde\delta \rho}{\rho} + \rho\nu -\tilde\tau\gamma.
\ee 
From \eqref{spinCoef}, we have $\nu = \tilde\delta V - V\tilde\tau$, and applying the vacuum condition $V = \frac{m}{2P^3}(\rho + \tilde\rho)$ from \eqref{Vvac}  leads to 
\be \label{nuEQ1}
\nu = -3V\frac{\tilde\delta P}{P} + \frac{\gamma}{\rho^2}\tilde\delta\rho -\frac{2\tilde\tau\gamma}{\rho}.
\ee 
Substituting \eqref{nuEQ1} into \eqref{Psi3EQ1}, and using \eqref{Vvac}, \eqref{gammaVac}, and the right-handed analog of \eqref{rhotauVac2} leads to 
\be \label{Psi3Final}
\Psi_3 = 3\gamma\Big(\,\frac{\tilde\delta\rho}{\rho} - 2\frac{\tilde\delta P}{P}\,\Big).
\ee 
On the other hand, from the expressions \eqref{NPMaxwell} for the Maxwell scalars and equation \eqref{Seq} for $S$ and again applying the vacuum condition $\gamma = \frac{m\rho^2}{2P^3}$ from \eqref{gammaVac}, we have 
\be \label{Psi3WeylRHS1}
\frac{1}{S}\varphi_1\varphi_2 = 3\rho\Big(\nu +V\frac{\tilde\delta P}{P} \Big).
\ee 
Inserting \eqref{nuEQ1} into \eqref{Psi3WeylRHS1}, and once again using \eqref{Vvac}, \eqref{gammaVac}, and the right-handed analog of \eqref{rhotauVac2} establishes that $\Psi_3 = \frac{1}{S}\varphi_1\varphi_2$, as in \eqref{NPWeyl}.

Since vacuum Kerr-Schild spacetimes are not generally of Petrov type D, one cannot expect the Weyl double copy to hold in general. As a consequence, we cannot assume that the last equation in \eqref{NPWeyl} for $\Psi_4$ will be automatically satisfied, unlike the equations for the other Weyl scalars. However, for any type D spacetime with a tetrad satisfying $\Psi_0 = \Psi_1 = 0$, the Weyl scalars must satisfy
\be\label{TypeDcondition}
3\Psi_2\Psi_4 = 2\Psi_3^2.
\ee
Putting $\Psi_2 = \frac{2}{3S}\varphi_1^2$ and $\Psi_3 = \frac{1}{S}\varphi_1\varphi_2$ into \eqref{TypeDcondition} immediately gives $\Psi_4 = \frac{1}{S}\varphi_2^2$. This completes the proof that our proposal \eqref{gauge} for the Kerr-Schild double copy implies the Weyl double copy when the spacetime is type D. In fact, what we have shown is somewhat stronger; since we have made no assumptions about the reality of the metric or about its signature, the left and right Weyl scalars are in general independent. The full Petrov type of such a spacetime is thus characterized by a pair of symbols $(\cdot, \tilde{\cdot})$, indicating the Petrov types of the left and right Weyl scalars. The argument we have made for the validity of the Weyl double copy applies separately to the left and right Weyl scalars, and so applies to complex spacetimes or real spacetimes in Kleinian signature with Petrov types $(D, \tilde{X})$ or $(X, \tilde{D})$, where $X \in \{I\hspace{-.25mm}I, D, O\}$. 
\section{Applications}\label{egs}
\subsection{Kerr}
Here we calculate the Kerr-Schild metric with Killing vector $X=(\partial_u+\epsilon\partial_v)/\sqrt{2}$ and $\Phi(Y)=-iaY$ with $a$ a constant. We see that $\epsilon=1$ corresponds to $X=\partial_t$ and $\epsilon=-1$ to $X=-\partial_z$. The roots of \eqref{fvac} are thus
\begin{equation}
    Y = \frac{v-\epsilon u-ia\pm\sqrt{(v-\epsilon u-ia)^2+4\epsilon\zeta\tilde\zeta}}{2\epsilon\tilde\zeta},
\end{equation}
and from here on out we choose to work with the $(-)$ solution. We find that
\be
    \rho = -2\left(v-\epsilon u-ia+\sqrt{(v-\epsilon u-ia)^2+4\epsilon\zeta\tilde\zeta}\right)^{-1}
\ee
and that $P = -\rho\sqrt{(v-\epsilon u-ia)^2+4\epsilon\zeta\tilde\zeta}$. With these one can now find the form of the metric. The Kerr metric can be obtained by letting $X$ be timelike, in which case $\epsilon=+1$. Its construction using the Newman-Penrose formalism for Kerr-Schild metrics can be read in \cite{mcintosh1988single}. Here we use the final result, namely that in coordinates $\{t,x,y,z\}$,
\begin{equation}
    l_\mu dx^\mu = \frac{\sqrt{2}r}{r+z}\Big[dt-\frac{rx-ay}{r^2+a^2}dx-\frac{ax+ry}{r^2+a^2}dy-\frac{z}{r}dz\Big]
\end{equation}
where $r$ is given implicitly by $(x^2+y^2)/(r^2+a^2)+z^2/r^2=1$, and that
\begin{equation}
    2V = -\frac{mr(r+z)^2}{r^4+a^2z^2}.
\end{equation}
Therefore the single copy is 
\begin{equation}
    A = h_{\mu\nu}X^\nu X^\mu = -\frac{2mr^3}{r^4+a^2z^2}\Big[dt-\frac{rx-ay}{r^2+a^2}dx-\frac{ax+ry}{r^2+a^2}dy-\frac{z}{r}dz\Big],
\end{equation}
which is, as expected, the result found via the standard stationary Kerr-Schild approach \cite{Monteiro:2014cda}.

\subsection{Schwarzschild and family}
With $a=0$, the Kerr-Schild graviton with a general signature Killing vector is
\begin{equation}
    h_{\mu\nu}dx^\mu dx^\nu = -\frac{2\sqrt{2}m}{\epsilon^2\sqrt{(v-\epsilon u)^2+4\epsilon\zeta\tilde\zeta}}\Bigg[\frac{1}{\sqrt{2}}d(\sqrt{(v-\epsilon u)^2+4\epsilon\zeta\tilde\zeta})-\frac{1}{\sqrt{2}}d(v+\epsilon u)\Bigg]^2.
\end{equation}
This metric is diffeomorphic to equation (28.21) of \cite{Stephani:2003tm}. For a timelike Killing vector, this solution becomes Schwarzschild. However, we find another solution that can be thought of as a cousin of Schwarzschild, one with a spacelike instead of a timelike Killing vector. Let $X=-\partial_z$ be the spacelike Killing vector. Therefore
\begin{equation}
    Y = \frac{-v-u-\sqrt{(v+u)^2-4\zeta\tilde\zeta}}{2\tilde\zeta}
\end{equation}
is the scalar function in null coordinates. The final metric can be written in a simpler way using cylindrical coordinates $\{t,\rho,\theta,z\}$. The new coordinates are $\rho=\sqrt{x^2+y^2}$ and $\theta = \arctan({y/x})$. The resulting metric is
\begin{equation}
    ds^2 = dt^2-d\rho^2-\rho^2d\theta^2-dz^2-\frac{2m}{\sqrt{t^2-\rho^2}}\Big[\frac{t}{\sqrt{t^2-\rho^2}}dt-\frac{\rho}{\sqrt{t^2-\rho^2}}d\rho+dz\Big]^2.
\end{equation}
Notice that this metric is cylindrically symmetric about the $z$-axis since
$\partial_z$ and $\partial_\theta = x\partial_y - y\partial_x$ are Killing vectors. There are also two boost Killing vectors that are most easily expressed in Cartesian coordinates as $\xi_x = x\partial_t + t\partial_x$ and $\xi_y = y\partial_t + t\partial_y$. Now the symmetry algebra satisfies the commutation relations $[\xi_x, \xi_y] = \partial_\theta$, $[\xi_x, \partial_\theta] = \xi_y$, and $[\xi_y, \partial_\theta] = -\xi_x$, with $X = -\partial_z$ being a central element that commutes with all others. Note that $X$ is also central in the Schwarzschild and Kerr cases. 

The single copy is
\begin{equation}
    A = \frac{q}{\sqrt{t^2-\rho^2}}\Big[\frac{t}{\sqrt{t^2-\rho^2}}dt-\frac{\rho}{\sqrt{t^2-\rho^2}}d\rho+dz\Big],
\end{equation}
which is gauge-equivalent to
\begin{equation}
    A' = \frac{q}{\sqrt{t^2-\rho^2}}dz.
\end{equation}
The zeroth copy is given by $\phi=h_{\mu\nu}X^\mu X^\nu = A_\mu X^\mu = q/\sqrt{t^2-\rho^2}$. This solution can be thought of as an infinitely long cylindrically symmetric singular shell that expands outward at the speed of light. The coordinates used above do not cover the region outside the shell (where the metric becomes complex). Inside the shell and around the $z$-axis, as $t$ increases, the components of the metric asymptotically approach their flat Minkowskian form, $\mathrm{diag}(1,-1,-\rho^2,-1)$.

When we set $\epsilon=0$, then $X=\partial_u$ and $P=1$. Since the $\epsilon\to 0$ limit is undefined in $h_{\mu\nu}$ above, we must start from scratch with the vacuum Kerr-Schild conditions. Doing so, we find that $Y = -v/\zeta$ and so $\rho = v/\zeta^2$. Thus $2V = mv(\zeta^2+\tilde\zeta^2)/(\zeta\tilde\zeta)^2$, and the Kerr-Schild graviton is
\begin{equation}
    h_{\mu\nu}dx^\mu dx^\nu = \frac{mv(\zeta^2+\tilde\zeta^2)}{(\zeta\tilde\zeta)^2}\Big[du - \frac{v}{\zeta\tilde\zeta}(\zeta d\zeta + \tilde\zeta d\tilde\zeta) + \frac{v^2}{\zeta\tilde\zeta}dv\Big]^2.
\end{equation}
The single copy is then
\begin{equation}
    A = \frac{mv(\zeta^2+\tilde\zeta^2)}{(\zeta\tilde\zeta)^2}\Big[du - \frac{v}{\zeta\tilde\zeta}(\zeta d\zeta + \tilde\zeta d\tilde\zeta) + \frac{v^2}{\zeta\tilde\zeta}dv\Big],
\end{equation}
and the zeroth copy is
\begin{equation}
    \phi = \frac{mv(\zeta^2+\tilde\zeta^2)}{(\zeta\tilde\zeta)^2}.
\end{equation}
This solution is a special case of a Kasner metric and, like the two examples above, diffeomorphic to a limit of equation (28.21)  in Section 28.1 of \cite{Stephani:2003tm}; here it is their $K=0$ limit.

\subsection{Self-dual Solutions}
A particularly interesting class of complex spacetimes are those that are \emph{self-dual}. In the Newman-Penrose formalism, self-duality is characterized by the vanishing of the right-handed Weyl scalars $\tilde\Psi_I = 0$. Of course, for real Lorentzian spacetimes, this implies that the left-handed Weyl scalars also vanish, since they are simply the complex conjugates of the right-handed Weyl scalars. In this case, the spacetime is conformally flat and the only possible self-dual solutions are maximally symmetric. However, as we have emphasized repeatedly, for complex spacetimes or real spacetimes with Kleinian signature, $\Psi_I$ and $\tilde\Psi_I$ are \it independent\rm, and non-trivial self-dual solutions are possible. Similarly, in electromagnetism, self-duality can be characterized by the vanishing of the right-handed Maxwell scalars $\tilde\varphi_i = 0$, and self-dual solutions are generically complex. Because the equations of motion are linear, real solutions of Maxwell's equations are in one-to-one correspondence with self-dual solutions. As one might expect, this is not the case for solutions of Einstein's equation, which are non-linear. However, it is well-known that for Kerr-Schild spacetimes with a geodesic Kerr-Schild vector, the Einstein equations linearize, so perhaps it is not surprising that there is a very simple correspondence between real vacuum Kerr-Schild spacetimes and self-dual spacetimes. Remarkably, the correspondence between real and self-dual vacuum Kerr-Schild spacetimes, as presented here, appears to have gone unnoticed in the existing literature until now.

In section \S \ref{VKS}, we found that vacuum Kerr-Schild spacetimes are characterized by a function $Y$ that is a solution to $f(Y, u + Y\tilde{\zeta}, \zeta + Yv) = 0$, where the function $f$ is given by \eqref{fvac}, and a second function $\tilde{Y}$ that satisfies the right-handed analog of the same equation. For real Lorentzian spacetimes, $\tilde{Y}$ is simply the complex conjugate of $Y$, so the solution is completely characterized by $Y$. However, for complex spacetimes, $Y$ and $\tilde{Y}$ are independent, and it turns out that the choice $\tilde{Y} = 0$ always leads to a self-dual solution of the vacuum Einstein equations.

To see that $\tilde{Y} = 0$ always leads to a self-dual solution, we focus on the right-handed spin coefficients. From the right-handed version of \eqref{spinCoef}, we can see immediately that $\tilde\rho$ and $\tilde\tau$ are zero, and since $\tilde\rho$ is zero, so is $\tilde\mu$. From the right-handed version of \eqref{gammaVac}, we also see that $\tilde\gamma = 0$. Finally, from the right-handed version of \eqref{spinCoef} we have $\tilde\nu = \delta V - V\tau$ and from \eqref{NPgroup1} $\delta\rho = \rho\tau$, which leads to $\tilde\nu = 0$ after using the vacuum condition \eqref{Vvac}. We conclude that for solutions with $\tilde{Y} = 0$, all of the right-handed spin coefficients vanish. Putting this result into the right-handed version of \eqref{NPgroup3} leads to $\tilde\Psi_I = 0$, so the spacetime is self-dual. It is simple to check that when $\tilde{Y} = 0$, the right handed version of the Maxwell scalars \eqref{NPMaxwell} also vanish, so the corresponding single copy is also self-dual.

\subsubsection{Self-dual analog of Kerr}
We illustrate the method described above to obtain the self-dual analog of the Kerr metric. One obtains the Kerr metric by setting the Killing vector $X=(\partial_u+\partial_v)/\sqrt{2}$ and setting $\Phi(Y) = -iaY$, where $a$ is the rotation parameter. $Y$ is thus a solution to a quadratic equation, and choosing the negative root gives
\begin{equation}
    Y = \frac{v-u-ia-\sqrt{(v-u-ia)^2+4\zeta\tilde\zeta}}{2\tilde\zeta}.
\end{equation}
Since $\tilde Y = 0$, we can write $\theta^2 = \mathrm{d}u+Y\mathrm{d}\tilde\zeta$. Furthermore the spin coefficient $\rho = \tilde\delta Y = \partial_\zeta Y - \tilde Y \partial_u Y = \partial_\zeta Y$, and $P = l_\mu X^\mu = 2^{-1/2}$. From these we compute the metric:
\begin{equation}
    ds^2 = 2dudv-2d\zeta d\tilde\zeta - \frac{2\sqrt{2}m}{\sqrt{(v-u-ia)^2+4\zeta\tilde\zeta}}\Big[du + \frac{v-u-ia-\sqrt{(v-u-ia)^2+4\zeta\tilde\zeta}}{2\tilde\zeta}d\tilde\zeta\Big]^2. \label{Self-Dual-Element}
\end{equation}
This is the self-dual analog of the Kerr solution obtained via setting $\tilde Y = 0$. A few comments are in order. First, we note that analytically continuing $a \to -ia$, and restricting to real values of all of the coordinates, results in a real metric with Kleinian signature. Furthermore, unlike in the case of the Kerr black hole, the rotation parameter does not correspond to a physical angular momentum in the self-dual analog, as it can be removed from the metric by performing the coordinate transformation $u' = u + ia$ or $v' = v - ia$ (before analytically continuing to imaginary $a$). The resulting metric can be expressed in Cartesian coordinates as 
\begin{equation}
    ds^2 = dt^2-dx^2+dy^2-dz^2-\frac{m}{r}\Big[dt-dz-\frac{r-z}{x-y}d(x-y)\Big]^2
\end{equation}
where $r^2\equiv x^2-y^2+z^2$.

This situation is reminiscent of the surprising discovery by Crawley, Guevara, Miller and Strominger that self-dual Kerr-Taub-NUT is diffeomorphic to Taub-NUT in Kleinian signature \cite{crawley2022black} (see also \cite{crawley2023self}). While we have not been able to find an explicit diffeomorphism taking one metric to the other, we suspect that the self-dual analog of Kerr we have presented here is in fact diffeomorphic to self-dual Taub-NUT. An analysis of a few quadratic, cubic and quartic curvature invariants (which we have included in appendix \ref{appendix B}) seems to support this conjecture. Establishing this conjecture would be an important result, since to our knowledge, it is not known that self-dual Taub-NUT is a single Kerr-Schild metric (although its double Kerr-Schild form is well known \cite{plebanski1976rotating,Chong:2004hw}). Conversely, if this indeed represents a newly discovered self-dual black hole, it would undoubtedly be of significant interest in its own regard. Additionally, it would indicate that the property of removing the rotation parameter through a diffeomorphism may persist more broadly in self-dual black hole spacetimes.  

\indent Returning to the double copy context, we obtain the single copy gauge field by contracting $h_{\mu\nu}$ with $X^\nu$:
\begin{equation}
    A = -\frac{2m}{\sqrt{(v-u-ia)^2+4\zeta\tilde\zeta}}\Big[du + \frac{v-u-a-\sqrt{(v-u-a)^2+4\zeta\tilde\zeta}}{2\tilde\zeta}d\tilde\zeta\Big].
\end{equation}
Similarly, the zeroth copy is
\begin{equation}
    \phi = -\frac{\sqrt{2}m}{\sqrt{(v-u-ia)^2+4\zeta\tilde\zeta}}.
\end{equation}
\subsubsection{The Eguchi-Hanson metric}
The Eguchi-Hanson metric is a self-dual, vacuum solution of Einstein's equation which is of the Kerr-Schild form and which has an expanding repeated principle null direction. This solution has been studied in the context of the double copy in \cite{Berman:2018hwd} and \cite{Luna:2018dpt}. Nevertheless, the Eguchi-Hanson metric does \emph{not} fall into the family of solutions we have studied in this article, which inlcudes all real Lorentzian vacuum Kerr-Schild spacetimes with an expanding Kerr-Schild vector, but only a subset of the complex and Kleinian signature spacetimes with the same properties. Although the Eguchi-Hanson metric lies outside the scope of the general results we have presented, we will find that examining the Killing vectors of this metric can still offer valuable insights and yield intriguing results.

The Eguchi Hanson metric is \cite{Eguchi:1978xp}
\be \label{Eguchi}
ds^2 = ds_0^2 + \frac{\lambda v^2}{(uv-\zeta\tilde\zeta)^3} \left(du - \frac{\zeta}{v} d\tilde\zeta\right)^2, 
\ee 
where $ds_0^2 = 2(dudv - d\zeta d\tilde\zeta)$, and $\lambda$ is a real parameter, which is neither a mass nor a cosmological constant. Taking all coordinates to be real results in the metric \eqref{Eguchi} in Kleinian signature. We may read off the metric functions
\be \label{EguchiFunctions}
Y = -\zeta/v, \qquad \tilde{Y} = 0, \qquad V = \frac{\lambda v^2}{2(uv - \zeta\tilde\zeta)^3}.  
\ee 
Now we can see why these metric functions do not define a metric of the type we have been considering: they do not obey the vacuum condition \eqref{Vvac}, which, given the above expressions for $Y$ and $\tilde{Y}$, should give $V = -\frac{m}{2v}$. To see this, note that $Y = -\zeta/v$ corresponds to putting $\Phi(Y) = 0$ and $X = \partial_u$ in \eqref{fvac}, thus $P = 1$, $\rho = -1/v$, and $\tilde{\rho} = 0$. Evidently, this construction does not lead to the Eguchi-Hanson metric, for which $\partial_u$ is not even a Killing vector. Rather, the Killing vectors for \eqref{Eguchi} may be written as
\bea
K_0 &=& u\partial_u - v\partial_v - \zeta\partial_\zeta + \tilde\zeta\partial_{\tilde\zeta} \nonumber \\
K_1 &=& \zeta\partial_v + u\partial_{\tilde\zeta} \nonumber \\ 
K_2 &=& \tilde\zeta\partial_u + v\partial_\zeta \nonumber \\
K_3 &=& u\partial_u - v\partial_v + \zeta\partial_\zeta - \tilde\zeta\partial_{\tilde\zeta} = \tilde{K}_0.
\eea 
Note that all of the $K_i$ above are also Killing vectors of the flat metric $ds_0^2$ and that $K_0$ is a central element of the symmetry algebra for the full metric \eqref{Eguchi}, i.e. $[K_0, K_i] = 0$. Similarly, the distinguished Killing vector $X$ was central in all of the examples we have studied. Furthermore, for the Eguchi-Hanson metric \eqref{Eguchi}, we can write 
\be \label{VwithQ}
V = -\frac{\lambda}{2Q^3}(\rho + \tilde\rho)
\ee 
where we have defined $Q \equiv l_\mu K_0^\mu = u - \zeta\tilde\zeta/v$ in analogy with $P = l_\mu X^\mu$. In view of this structure, it is tempting to consider both the `standard' gauge field $A_\mu = h_{\mu\nu}X^\nu$ and the modified gauge field $A^{\textrm{mod}}_\mu = h_{\mu\nu}K_0^\nu$. For the standard gauge field, we have
\be 
A_\mu = \frac{\lambda v^2}{(uv - \zeta\tilde\zeta)^3}l_\mu,
\ee 
while for the modified gauge field we have
\be 
A^{\textrm{mod}}_\mu = \frac{\lambda v}{(uv - \zeta\tilde\zeta)^2}l_\mu,
\ee 
both of which are self-dual solutions of the flat-space vacuum Maxwell equations!\footnote{And neither of which agrees with the self-dual single copy in \cite{Berman:2018hwd}.}

We may also consider a real Lorentzian analog of the Eguchi-Hanson metric; however, since the Eguchi-Hanson metric is not contained in the family of solutions we have considered in our general discussion, there is no guarantee that the real analog will be a solution of the vacuum Einstein equations. Indeed, since the family we have considered in this article contains all real Lorentzian vacuum Kerr-Schild solutions with an expanding repeated principle null direction, one would expect that the real Lorentzian analog of the Eguchi-Hanson metric must not be a vacuum solution. Let us consider the line element
\be \label{EH-real}
ds^2 = ds_0^2 + \tfrac{\lambda v^2}{(uv - \zeta\tilde\zeta)^3}\left(du - \tfrac{\tilde\zeta}{v}d\zeta - \tfrac{\zeta}{v}d\tilde\zeta + \tfrac{\zeta\tilde\zeta}{v^2}dv\right)^2,
\ee 
which corresponds to a pure-radiation stress tensor 
\be 
T_{\mu\nu} = \frac{24\pi G\lambda}{(uv - \zeta\tilde\zeta)^4}l_\mu l_\nu.
\ee 
The line element \eqref{EH-real} has metric functions 
\be \label{EguchiFunctions}
Y = -\zeta/v, \qquad \tilde{Y} = -\tilde\zeta/v, \qquad V = \frac{\lambda v^2}{(uv - \zeta\tilde\zeta)^3}, 
\ee 
where once again $V$ satisfies \eqref{VwithQ} because the new terms that have been introduced in $l_\mu$ to make the line element real are orthogonal to $K_0^\mu$. The line element \eqref{EH-real} has four Killing vectors, three of which are also Killing vectors of the self-dual metric \eqref{Eguchi}, namely $K'_0 = K_0$, $K'_1 = \zeta\partial_u + v\partial_{\tilde\zeta} = \tilde{K}_2$, $K'_2 = K_2$, $K'_3 = K_3 = \tilde{K}_0$. A few points are worth noting here. First, $K_0 = K'_0$ remains a Killing vector of the new metric \eqref{EH-real}, but it is no longer central, as it does not commute with $K'_1$. Second, even though $K_0$ is complex, its imaginary part is orthogonal to $l_\mu$, so it can still be used to construct a \emph{real} gauge field $A^{\textrm{mod}}_\mu = h_{\mu\nu}K_0^\nu$ in addition to the standard gauge field $A_\mu = h_{\mu\nu}X^\nu$ with $X = \partial_u$. We have
\be \label{EH-real-gauge-standard}
A_\mu = \frac{2\lambda v^2}{(uv - \zeta\tilde\zeta)^3}l_\mu,
\ee 
and
\be \label{EH-real-gauge-mod}
A^{\textrm{mod}}_\mu = \frac{2\lambda v}{(uv - \zeta\tilde\zeta)^2}l_\mu.
\ee 
As one might expect, these are not solutions of the vacuum Maxwell equations, but rather correspond to sources $J_\mu = \nabla^{(0)\nu} F_{\mu\nu}$ and $J^{\textrm{mod}}_\mu = \nabla^{(0)\nu} F^{\textrm{mod}}_{\mu\nu}$ given by
\be 
J_\mu = \frac{6\lambda v^2}{(uv -\zeta\tilde\zeta)^4}l_\mu,
\ee 
and 
\be 
J^{\textrm{mod}}_\mu = \frac{4\lambda v}{(uv -\zeta\tilde\zeta)^3}l_\mu.
\ee 
In either case, we see that the double copy for fields leads to a double copy for sources of the form $T_{\mu\nu} \propto J_\mu J_\nu \propto J^{\textrm{mod}}_\mu J^{\textrm{mod}}_\nu$, which has been observed in the double copy analysis of Kinnersley's photon rocket spacetime \cite{Kinnersley:1969zz} which was carried out in \cite{Luna:2016due}.

\subsection{A simple solution with Petrov type II}

The Petrov type of the solution is determined by the arbitrary function $\Phi(Y)$, with the solution having Petrov type II if and only if $\Phi'''(Y) \neq 0$ \cite{mcintosh1988single}. One can see why simple examples of closed form Kerr-Schild solutions with Petrov type II are in short supply: one needs to solve equation \eqref{fvac} for some choice of $\Phi(Y)$ with $\Phi'''(Y) \neq 0$, but the latter requirement typically makes the equation of cubic order or higher; or worse, requires us to solve a non-polynomial equation for $Y$. However, there is a simple case where the equation can be solved explicitly and exactly. First notice that when $X = \partial_u$, the equation for $Y$ becomes
 \be
\Phi(Y) + \zeta + vY = 0. \label{Kerr Theorem, null case}
\ee 
Next we make the choice
\be
\Phi(Y) = a\sqrt{Y} \label{Q choice}
\ee
where $a$ is a real constant with dimensions of length. With this choice, equation \eqref{Kerr Theorem, null case} becomes a quadratic equation for $\sqrt{Y}$, which leads to the solution
\be
Y = \left(\frac{a}{2v}\right)^2\left(1 \pm \sqrt{1 - 4\frac{v\zeta}{a^2}}\right)^2.
\ee
 Now using equation \eqref{Vvac}, we find
\be
V = \frac{m}{2v} + \frac{ma}{4v}\left(\frac{1}{\sqrt{a^2 - 4v\zeta}} + \frac{1}{\sqrt{a^2 - 4v\tilde\zeta}}\right).
\ee
The spin coefficients are:
\begin{gather}
\rho = -\frac{1}{v}\left(1 + \frac{a}{\sqrt{{a}^{2}-4\,v\zeta}}\right) \\
\gamma = \frac{m\rho^2}{2} \\
\mu = \rho V \\
\nu = -{\frac {ma}{2} \left( {a}^{2}-4\,v\zeta \right) ^{-{\frac{3}{2}}}} \\
\kappa = \sigma = \alpha = \beta = \tau = \lambda = \epsilon = 0.
\end{gather}

The Weyl scalars are:
\begin{gather}
\Psi_0 = \Psi_1 = 0 \\
\Psi_2 = -\frac{m}{{v}^{3}}\,{\frac { \left( a+\sqrt {{a}^{2}-4\,v\zeta} \right) ^{3}}{ \left( {a
}^{2}-4\,v\zeta \right) ^{3/2}}} \\
\Psi_3 = \frac{3ma}{v}{\frac {\left( a+\sqrt {{a}^{2}-4\,v\zeta} \right) }{ \left( {a}^{2
}-4\,v\zeta \right) ^{2}}} \\
\Psi_4 = -{\frac {6mav}{ \left( {a}^{2}-4\,v\zeta \right) ^{5/2}}}.
\end{gather}
It can be checked that $2\Psi_3^2 - 3\Psi_2\Psi_4 \neq 0$, so the solution is indeed of Petrov type II. Since we have taken $X = \partial_u$, we have $P = 1$, and the single copy is simply $A_\mu = 2Vl_\mu$, which can be checked to solve the flat space vacuum Maxwell equations. The zeroth copy is simply $\phi = V$, which clearly solves the vacuum wave equation as $\partial_u\partial_v V$ = $\partial_\zeta\partial_{\tilde{\zeta}} V$ = 0. 

\section{Discussion}
We have shown that infinitesimal isometries play a crucial role in the Kerr-Schild double copy construction. In the context of four-dimensional vacuum spacetimes, the geodesic and shear-free character of the Kerr-Schild vector and the existence of a distinguished Killing vector allow us to define a classical double copy without making any additional assumptions. In particular, it can be defined even when the Killing vector is null or spacelike with respect to the background metric, allowing us to dispense with the restriction that the spacetime be stationary. In hindsight, there are numerous reasons to be suspicious of this requirement. Even in the simplest example of the Schwarzschild black hole, the spacetime is only stationary outside the event horizon. Meanwhile, the Kerr spacetime contains a stationary region with closed timelike curves where the timelike Killing vector tangent to these curves does \emph{not} lead to a solution of Maxwell's equations via contraction with the graviton. This state of affairs might lead one to speculate that the Killing vector must be timelike with respect to the background metric, however our results clearly indicate otherwise. 

Since we have fixed a null tetrad suitable for vacuum Kerr-Schild spacetimes from the outset and made use of the spin coefficient formalism of Newman and Penrose, our approach is particularly well-suited to make contact with the Weyl double copy. Although the Weyl double copy is usually formulated in spinorial language, it admits a very simple rewriting in terms of the Newman-Penrose Weyl scalars and Maxwell scalars, allowing us to prove that the reformulation of the Kerr-Schild double copy we have proposed implies the Weyl double copy when the spacetime is Petrov type D. Interestingly, four out of the five complex scalar equations encoding the Weyl double copy in Newman-Penrose language are satisfied without making any assumption about the Petrov type of the spacetime, and the type D condition is only required in order to satisfy the component of the Weyl double copy corresponding to $\Psi_4$, which is associated with radiation. This suggests that it may be possible to formulate an exact Weyl double copy for Kerr-Schild spacetimes of any Petrov type by including an additional radiation term along the lines of 
$\Psi_{ABCD} = \frac{1}{S_{\tiny{\textrm{Coul}}}}\varphi^{\tiny{\textrm{Coul}}}_{(AB}\varphi^{\tiny{\textrm{Coul}}}_{CD)} + \frac{1}{S_{\tiny{\textrm{Rad}}}}\varphi^{\tiny{\textrm{Rad}}}_{(AB}\varphi^{\tiny{\textrm{Rad}}}_{CD)}$, where the Coulombic part is the standard Weyl double copy and the radiation part vanishes for type D spacetimes. Such a scheme is also reminiscent of recent explorations into the Weyl double copy with sources \cite{Easson:2021asd,Easson:2022zoh}. In terms of the Newman-Penrose Maxwell scalars, this would correspond to radiation field with $\varphi^{\tiny{\textrm{Rad}}}_0 = \varphi^{\tiny{\textrm{Rad}}}_1 = 0$. Indeed, the twistorial approach to the classical double copy \cite{White:2020sfn, Chacon:2021wbr} can be applied to solutions of \emph{linearized} gravity with any Petrov type, while vacuum Kerr-Schild spacetimes are simultaneously solutions of linearized gravity and the full non-linear Einstein equations. It stands to reason that the twistorial double copy should imply an exact Weyl double copy for Kerr-Schild spacetimes, and our results hint at what this exact relation may look like.

We have presented a number of examples of vacuum Kerr-Schild spacetimes and their single copy gauge fields that have not previously been studied in this context. Perhaps the most interesting example we have studied is the self-dual analog of the Kerr spacetime in Kleinian signature. While we have speculated that this spacetime is diffeomorphic to the Kleinian self-dual Taub-NUT metric, we have not been successful in finding an explicit diffeomorphism that confirms this conjecture. Either confirming or refuting this conjecture would be an important result, as the former would establish that Kleinian self-dual Taub-NUT is a single Kerr-Schild spacetime, while the latter would indicate that the self-dual analog of Kerr we have presented here is in fact a genuinely new self-dual black hole spacetime. 

Our work suggests a number of straightforward extensions. Developing a deeper understanding of the map between real Lorentzian and self-dual vacuum Kerr-Schild spacetimes in general and the  self-dual analog of Kerr in particular would be worthwhile endeavors. The self-dual setting also offers the possibility of developing a non-abelian classical double copy via the close analogy between Penrose's non-linear graviton proposal \cite{penrose1976nonlinear} and the Penrose-Ward correspondence for Yang-Mills instantons \cite{atiyah1977instantons}. Pursuing this approach offers the possibility of making closer contact between the classical and quantum versions of the double copy, and deepening our understanding of the role of twistor theory in the double copy paradigm. Additionally, there exist some simple generalizations of the null tetrad we have used that accommodate solutions of Einstein-Maxwell theory as well as spacetimes with a pure radiation stress tensor. In the case of Kerr-Schild spacetimes with pure radiation stress tensors, examples of interest, including the real analog of the Eguchi-Hanson metric presented here and Kinnersley's photon rocket, exhibit a double copy for sources that mimics the double copy structure of the fields---it would be interesting to see whether this feature persists in a more systematic analysis of such non-vacuum spacetimes. Since we have made use of Newman-Penrose formalism, our results are necessarily restricted to four-dimensional spacetimes. It would be a major improvement on our work if at least some of the results could be extended to spacetimes of arbitrary dimension. Finally, the tetrad we have used here is the same one used in the study of the so-called Newman-Penrose map \cite{Elor:2020nqe, Farnsworth:2021wvs}. It would be interesting to use the results of this work to test the degree to which the Newman-Penrose map is compatible with the Kerr-Schild double copy in a general setting.

\section{Acknowledgements}
We are grateful to Kara Farnsworth, Michael Graesser, and Cindy Keeler for helpful discussions. DAE is supported in part by the U.S. Department of Energy, Office of High Energy Physics, under Award Number DE-SC0019470. TM  is  supported  by  the Simons Foundation, Award 896696.

\appendix
\section{Calculation of $\Delta\rho$}\label{SomeCalcs}
We have the Bianchi identity
\be  
\delta\Psi_2 = 3\tau\Psi_2 \label{Bianchi}
\ee  
along with the following expression for $\Psi_2$ which is valid in vacuum:
\be  \label{Psi2vac}
\Psi_2 = \mu\rho + \gamma(\rho - \tilde\rho).
\ee  
Acting with $\delta$ on \eqref{Psi2vac} gives 
\bea
\delta\Psi_2 &=& \rho\delta\mu + \mu\delta\rho + \delta\gamma(\rho -\tilde\rho) + \gamma\delta(\rho -\tilde\rho) \nonumber \\
&=& \rho\delta(V\rho) + (2\mu + \gamma)(\rho - \tilde\rho)\tau + \gamma\delta\rho - \gamma\delta\tilde\rho \nonumber \\
&=& \rho^2\delta V + \rho V\delta\rho + (2\mu + \gamma)(\rho - \tilde\rho)\tau + \gamma\delta\rho - \gamma\delta\tilde\rho \nonumber \\
&=&  \rho^2(\tilde\nu + V\tau) + \rho V(\rho -\tilde\rho)\tau + (2\mu + \gamma)(\rho - \tilde\rho)\tau + \gamma(\rho-\tilde\rho)\tau - \gamma\delta\tilde\rho \nonumber \\
&=&  \rho^2(\tilde\nu + V\tau) + (3\mu + 2\gamma)(\rho - \tilde\rho)\tau - \gamma\delta\tilde\rho \nonumber \\
\label{deltaPhi2-1}
\eea 
where we used $\mu = \rho V$, $\tilde\nu = \delta V - V\tau$ from \eqref{spinCoef}, and the (vacuum) Newman-Penrose equations $\delta\rho = (\rho - \tilde\rho)\tau$ from \eqref{NPgroup1} and $\delta\gamma = (\mu + \gamma)\tau$ from \eqref{NPgroup2}.
Meanwhile, putting \eqref{Psi2vac} into the right hand side of \eqref{Bianchi} gives
\bea
\delta\Psi_2 &=& 3\tau\mu\rho + 3\tau\gamma(\rho -\tilde\rho) \nonumber \\
&=& 3\tau(\mu + \gamma)(\rho -\tilde\rho) + 3\tau\mu\tilde\rho.\label{deltaPhi2-2}
\eea  
Combining \eqref{deltaPhi2-1} and \eqref{deltaPhi2-2} leaves us with an expression for $\delta\tilde\rho$ which is not found in the Newman-Penrose equations:
\be 
\delta\tilde\rho = \frac{1}{\gamma}\Big(\rho^2(\
\tilde\nu + V\tau) - 3\tau\mu\tilde\rho - \gamma(\rho -\tilde\rho)\tau\Big). \label{deltaRhoTildeEq}
\ee
Now from \eqref{NPgroup1}, we have the (vacuum) Newman-Penrose equation 
\be \label{DeltaRhoNP}  
\Delta\rho = \tilde\delta\tau - \tau\tilde\tau - V\rho^2 
\ee
from which we can isolate an expression for $\Delta\rho$. First, we recall that 
\be 
\tau = -\frac{\rho}{P}(bY + c).
\ee 
Acting on the equation above with $\tilde\delta$ leads to 
\bea
\tilde\delta\tau &=& \tilde\delta\left(-\frac{\rho}{P}(bY + c)\right) \nonumber \\
&=& -\frac{\tilde\delta\rho}{P}(bY + c) + \frac{\rho\delta P}{P^2}(bY + c) - b\frac{\rho^2}{P} \nonumber \\
&=& -\frac{1}{P}(bY + c)\frac{1}{\tilde\gamma}\left[\tilde\rho^2(\nu + V\tilde\tau) - 3\tilde\tau\tilde\mu\rho + \tilde\gamma(\rho -\tilde\rho)\tilde\tau\right] + \frac{\rho^2}{P^2}|bY+ c|^2 - b\frac{\rho^2}{P}.
\eea 
Substituting this last equation into \eqref{DeltaRhoNP} then gives us
\bea 
\Delta\rho 
&=& 
-\frac{1}{P}(bY + c)\left[\frac{2P^3}{m}\nu + 2\rho\tilde\tau + \frac{3}{P}(b\tilde{Y} + \tilde{c})\rho(\rho + \tilde\rho)\right] + \frac{\rho^2}{P^2}|bY+ c|^2 - b\frac{\rho^2}{P} -\tau\tilde\tau - V\rho^2. \nonumber \\ \label{DeltaRhoCalc}
\eea  

\section{Curvature invariants for self-dual black holes}\label{appendix B}
Quadratic, cubic and quartic curvature invariants for the self-dual analog of Kerr \eqref{Self-Dual-Element} can be computed straightforwardly with the help of an appropriate symbolic algebra system such as Maple's differential geometry package, or Mathematica's xAct suite. The results are summarized below:
\begin{eqnarray}
R_{\m\n\s\r}R^{\m\n\s\r} &=& \frac{192\, m^2}{R^6} \\
R_{\m\n}{}^{\a\b}R_{\a\b}{}^{\r\s}R_{\r\s}{}^{\m\n} &=& \frac{768 \sqrt{2} m^3}{R^9} \\
R_{\alpha\beta\gamma\rho} R^{\alpha\beta\gamma}{}_{\sigma} R_{\mu\nu\kappa}{}^{\rho} R^{\mu\nu\kappa\sigma} &=& \frac{9216 m^4}{R^{12}} \,,
%
\end{eqnarray}
where we have defined $R^2 = u^2 +v^2 -2uv + 4\zeta\tilde\zeta$. \\
\indent The metric for the self-dual Taub-NUT spacetime can be written in the form \cite{crawley2022black}:
\be \label{self-dual-Taub-NUT}
ds^2 = \frac{r-m}{r+m}(dt - 2M\cosh\theta \,d\phi)^2 + \frac{r+m}{r-m}\,dr^2 - (r^2 - m^2)(d\theta^2 + \sinh^2\theta\, d\phi^2).
\ee 
The corresponding curvature invariants for the self-dual Taub-NUT metric are:
\begin{eqnarray}
R_{\m\n\s\r}R^{\m\n\s\r} &=& \frac{96\, M^2}{(r+M)^6} \\
R_{\m\n}{}^{\a\b}R_{\a\b}{}^{\r\s}R_{\r\s}{}^{\m\n} &=& \frac{384 M^3}{(r+M)^9}\\
R_{\alpha\beta\gamma\rho} R^{\alpha\beta\gamma}{}_{\sigma} R_{\mu\nu\kappa}{}^{\rho} R^{\mu\nu\kappa\sigma} &=& \frac{2304 M^4}{(r+M)^{12}}.
\end{eqnarray}
The above invariants match for the two metrics under the identification
\be\label{rs2}
R \rightarrow r + M \,, \qquad m \rightarrow \frac{M}{ \sqrt{2}} 
\ee
Furthermore, the self-dual black hole \eqref{Self-Dual-Element} only has two obvious Killing vectors, yet the above curvature invariants depend only on a radial coordinate, suggesting that the solution may indeed be fully spherically symmetric as is the self-dual Taub-NUT metric. Even still, finding an explicit diffeomorphism between \eqref{Self-Dual-Element} and \eqref{self-dual-Taub-NUT} has proven to be surprisingly subtle, and we leave a definitive resolution of this problem for future work.

\bibliographystyle{JHEP}
\bibliography{bib}

\end{document}